\documentclass[12pt]{iopart}
\usepackage{amsmath,bm,amssymb}
\usepackage{mathrsfs}
\usepackage{amsfonts}
\usepackage[dvips]{graphicx}
\usepackage{dcolumn}
\newcommand{\ind}{{\mathbf kn}}
\newcommand{\beq}{\begin{equation}}
\newcommand{\eeq}{\end{equation}}

\newcommand{\half}{\frac{1}{2}}

\newcommand{\lbrla}{\left\langle }
\newcommand{\rbrra}{\right\rangle }

\newcommand{\lbrp}{\left| }
\newcommand{\rbrp}{\right| }

\newcommand{\bra}[1]{\lbrla #1 \rbrp}
\newcommand{\ket}[1]{\lbrp #1 \rbrra}
\newcommand{\braket}[2]{\lbrla #1 | #2 \rbrra}

\newcommand{\ignore}[1]{}

\newcommand{\D}{\text{d}}
\newcommand{\ks}{{\mathbf k\sigma}}
\newcommand{\eligand}{\epsilon_\ks^{\phantom\dagger}}

\begin{document}

\title[The Kondo signature of CuPc on Ag(100)]%
  {Multi-orbital Non-Crossing Approximation from maximally localized
  Wannier functions: 
  the Kondo signature of copper phthalocyanine on Ag (100) }
\author{Richard Koryt\'ar$^1$}
  \ead{rkorytar@cin2.es}
\author{Nicol\'as Lorente$^1$}
\address{$^1$Centro de investigaci\'on en
  nanociencia y nanotecnolog\'{\i}a
  (CSIC - ICN), Campus de la UAB, 
  E-08193 Bellaterra, Spain}
\date{\today}
\begin{abstract}
We have developed a multi-orbital approach to compute the electronic structure
of a quantum impurity using the non-crossing approximation. The calculation
starts with a mean-field evaluation of the system's electronic structure using
a standard quantum chemistry code. Here we use density functional theory (DFT).
We transformed the one-electron structure into an impurity Hamiltonian by using
maximally localized Wannier functions (MLWF). 
Hence, we have developed a method
to study the Kondo effect in systems based on an
initial one-electron calculation. We have applied our methodology to a copper
phthalocyanine molecule chemisorbed on Ag (100), and we have described its
spectral function for three different cases where the molecule presents a
single spin or two spins with ferro- and anti-ferromagnetic exchange couplings. 
We find that the use of broken-symmetry
mean-field theories such as Kohn-Sham DFT cannot deal with the complexity
of the spin of open-shell molecules on metal surfaces and extra modeling
is needed.  
\end{abstract} 
\maketitle

\section{Introduction}

Since the study of the Kondo features of Ce ad-atoms on Ag
(111)\cite{Berndt:PRL} and Co ad-atoms on Au (111) \cite{Crommie:Science},
 the scanning tunnelling microscope (STM) has become a privileged
tool in the study of surface Kondo physics. The STM is a non-intrusive
probe that can address adsorbed objects at very low bias with very small
currents. Hence, the STM basically explores the equilibrium properties
of the adsorbed systems. Besides ad-atoms, Kondo
physics has been revealed in objects of increasing
complexity, from single adsorbates \cite{Berndt:PRL,Crommie:Science,Knorr:PRL}
to large organic molecules \cite{Zhao:Science,Iancu:PRL,Gao:PRL,Pascual:PRL,Hla:PRL},
ordered nanostructures \cite{Wegner:PRL} and ad-atom-ligand
structures \cite{Wahl:PRL,Liechtenstein:PRL}.

Several recent reports show that organic molecules display a Kondo state due to
a spin in their extended $\pi$-orbitals even when they are adsorbed on a metal
surface \cite{Pascual:PRL,Hla:PRL,Komeda:Nat,Mugarza:Science}. This is somewhat
of a surprise for two reasons: $(i)$ some of these molecules are magnetic in
the gas phase because they have a magnetic atom, $(ii)$ an extended
$\pi$-orbital is expected to have a large  overlap with the metal surface
likely driving the $\pi$-system into a mixed valence regime instead of a Kondo
one.  Hence, the appearance of Kondo physics in these systems depends on a
series of parameters where common wisdom is likely to fail.  It is then of
great interest to perform calculations to rationalize the particular features
of adsorbed large molecules that depend as little as possible on
adjustable parameters. 

We have implemented an impurity solver for the Anderson
Hamiltonian \cite{Anderson:PR} that reads the one-electron structure from a
density functional theory (DFT) calculation for a given spin configuration of
the impurity (in the present case the molecular orbitals involved in the spin
configuration), uses the DFT hybridization to compute the dynamical electron
exchange with the substrate, and assumes a single electron fluctuation which
corresponds to the $U\rightarrow \infty$ limit of the Anderson Hamiltonian. The
impurity solver uses the non-crossing approximation
NCA \cite{GreweNCA,Kuramoto:ZPB,Coleman:PRB,Bickers:RMP} in the multi-orbital formalism
by Kuramoto \cite{Kuramoto:ZPB}. The multi-orbital aspects of the Kondo
problem must be correctly taken into account in realistic accounts of the Kondo
problem as shown by Kroha and collaborators \cite{Kroha:PRL,Kroha:PhysE}.

The NCA is a reliable approximation \cite{Grewe} away from very low
temperatures where it fails to reproduce the Fermi-liquid behavior in
fully screened Kondo systems \cite{Kroha:JPN} and where spurious spectral feature
appears at the Fermi energy \cite{Kroha:JPN,Kuramoto3:ZP}.  At
typical experimental temperatures NCA is a good choice for this type of
calculation because it retains all the electronic structure of the one-electron
part of the Hamiltonian while keeping a correct description of the main Kondo
features. In fact, the main limitation of our theory rather comes from
the use of the customary local or semi-local approximations to DFT.  Indeed,
DFT calculations on electronic gaps lead to discrepancies of a
factor 2 off the experimental gap \cite{Godby:prl}. This is
definitely a big drawback when evaluating Kondo temperatures ($T_K$) because they
depend exponentially on the molecular level value. The Kondo temperature
depends on the ratio of the molecular eigenvalue to its broadening,  
due to the molecular hybridization with the
metal substrate. We will use this ratio as a parameter.
Yet, our procedure retains the symmetry and
the relative strengths of the one-electron Hamiltonian as given by DFT. A
second important approximation of our work is that of the infinite intramolecular
Coulomb energy, $U\rightarrow \infty$. Extensions of
NCA to treat finite $U$ \cite{Haule,Kuramoto,Grewe:EPJB}
lead to considerable improvement of the $T_K$. However, since
the LDA based electronic structure impedes the calculation of the
$T_K$, the infinite $U$ approximation largely suffices
for our purposes.

In NCA, The Kondo physics is modeled by virtual fluctuations of the impurity
occupancy. These fluctuations are made
possible by coupling the Kondo impurity to the substrate.
In principle, NCA can treat all possible
configurations. The
$U\rightarrow \infty$ approximation is admissible, provided
that the self-energy contribution due to configurations with
two electrons or more is negligible. One can generalize
this idea and indeed consider two sets of configurations 
which differ by the addition of one electron, energetically 
separated from other configurations by some large value ``$U$''
 \cite{Kuramoto:ZPB}.
Hence, with NCA we can treat impurities of increasing
complexity, where the physics involved will correspond to 
virtual fluctuations among configurations differing in electron
number by one.

In this way, Roura Bas and Aligia have treated the singlet-triplet quantum
transition of an Anderson impurity within NCA \cite{Roura:PRB,Roura:JPCM}.  We
use a similar approach to describe the configurations of a copper
phthalocyanine (CuPc) on Ag (100). On this surface, CuPc captures one electron
from the substrate \cite{Mugarza:prl} while maintaining a very localized spin on
the copper atom at the center of the molecule.  Hence, this system is properly
characterized  by a Hamiltonian describing singlet-triplet transitions.
In order to perform our multi-orbital NCA
calculations, we first simulate CuPc/Ag(100) with DFT, we transform to a
maximally-localized Wannier function (MLWF) basis set \cite{Korytar:JPCM}, and
solve the DFT Hamiltonian expressed in this basis set with our multi-orbital
NCA code, selecting molecular orbitals that take part in the Kondo physics.

The use of MLWF is mandatory to be able to unambiguously transform the DFT
Hamiltonian into an Anderson-like one as we have shown in
\cite{Korytar:JPCM}. This is perhaps the biggest difference with
other works using impurity solvers based on DFT
calculations \cite{Gorelov,Kotliar,Dagotto}. Our approach is thus algorithmic,
except for the tuning of the molecular-orbital levels with respect to their
hybridization with the substrate since these are quantities where current DFT
approaches fail.

In the following section we give more details regarding the implementation and
execution of the calculations on CuPc/Ag(100).  In section~\ref{Results}, we
first show the DFT results and their conversion to the MLWF basis set. The MLWF
results permit us to explore the electronic structure that is involved in Kondo
physics. We also evaluate the spectral functions for the lowest unoccupied
molecular orbital (LUMO) that is actively involved in the generation of Kondo
spectral features. Finally, we analyze our results and conclude this work.

\section{Method}

The NCA is particularly adequate for describing the electronic
structure at different energy scales. Our multi-orbital
implementation \cite{Kuramoto:ZPB} uses the mean-field results of a local
density approximation (LDA) calculation, transforms the LDA Hamiltonian
into a MLWF basis which permits us to write an Anderson Hamiltonian,
$U\rightarrow\infty$, while keeping the full multi-configurational aspects
of the problem \cite{Korytar:JPCM}. The NCA is applied on the obtained
Anderson Hamiltonian. In this way, the calculated electronic structure
contains all molecular+substrate information in the presence of Kondo
physics. In this section, we give details on how this is achieved,
with special care in the choice of the configurations that will determine
the Kondo physics of CuPc on Ag (100).

\subsection{Density functional calculations}

Density functional calculations of gas phase copper phthalocyanine (CuPc)
capture the relevant electronic and geometric properties of the
molecule \cite{Liao:JCP}. Briefly, CuPc is a D$_{4h}$ molecule,
\fref{CuPc}, where the $d$ electrons of the central copper atom are split
by the ligand field of the surrounding atoms. The ligands capture $\sim 2$
electrons of the Cu atom rendering the molecule in a magnetic $d^9$
configuration \cite{Liao:JCP}. The $d$ manifold
is split such that the $d_{x^2-y^2}$ orbital is singly occupied (SOMO, the
singly occupied molecular orbital).  The following empty orbital (LUMO, lowest
unoccupied molecular orbital) has $\pi$ character and double degeneracy,
because it constitutes the $e_g$ representation \cite{Liao:JCP} of the point
group. 
 
Previous studies using LDA of CuPc on Ag (100) show that the main electronic
properties of the adsorbed molecule are retained in the
calculation \cite{Mugarza:prl}. Here, we take on that work and extract the
relevant one-electron physics important for determining the Kondo state. 

We have used the \textsc{Siesta} code \cite{Soler:JPCM}, relaxed the molecule
and first two surface layers to forces below 0.04 eV/\AA~using the geometry of
the reference~\cite{Mugarza:prl}.  The calculations are periodic, using a
super-cell containing five layers of $7\times 7$ Ag atoms. This unit
cell size is converged for computing electronic structure features
of the adsorbed CuPc on Ag(100) \cite{Mugarza:prl}. The electronic
structure calculations have been done with norm-conserving
pseudo-potentials \cite{Troullier:PRB} and strictly localized, DZP numerical
atom-centered basis set, optimized for this system \cite{Mugarza:prl}. The
basis sets are build using the method by Anglada \emph{et
al.} \cite{Anglada:PRB}, and we have used the improved noble metal surface
description using the basis sets of the reference \cite{Sandra:PRB}. The
confinement radii (all in Bohr) of s-type first-$\zeta$ basis functions
were 6.06 for hydrogen, 5.20 (C), 5.64 (N), 5.77 (Cu) and 5.55 (Ag). The
confinement radius of the first-$\zeta$ extended diffuse orbitals was 7.75.
Good agreement with our previous work \cite{Mugarza:prl} is thus attained.

The Kohn-Sham Bloch functions were calculated on a Monkhorst-Pack grid
$2\times2\times 1$ and transformed to the basis of maximally localized Wannier
functions \cite{Marzari:PRB,Souza:PRB,Mostofi:CPC} (MLWF).  Our method for
obtaining MLWF from \textsc{Siesta} is described in
\cite{Korytar:JPCM}.

\subsection{Maximally-localized Wannier functions}

The main reasons for working with a MLWF basis set are: $(i)$ 
the reduction of the computational problem by projecting
the $M$-dimensional Hilbert space of Kohn-Sham states, where $M$ is the
full dimension of the CuPc plus Ag (100) electronic problem, onto a smaller
$N$-dimensional space which faithfully represents \cite{Souza:PRB} the energy
bands around the Fermi level taking part in the Kondo effect; $(ii)$ 
the orthogonality of the MLWF basis set that permits us to use standard 
multi-orbital NCA \cite{Kuramoto:ZPB,Kroha:JPN}, and $(iii)$ the extreme
localization of MLWF plus their orthogonality gives us a natural way to
partition the problem into impurity and substrate 
subspaces \cite{Korytar:JPCM}.

Maximally localized Wannier functions have been used in the study of
strongly correlated matter 
(for instance see \cite{Sakakibara:PRL,Ikeda:PRB,Lechermann:PRB,Karolak:JPCM}). In some
cases, the spread minimization is skipped and the Kohn-Sham electronic
structure is projected directly onto trial orbitals. This strategy has
been used in the study of transition metal oxides~\cite{Korotin:EPJ}.
In our system, we found that the projection onto trial
orbitals gives unreliable results. The disentanglement method~\cite{Souza:PRB} 
can be contrasted with direct selection of certain bands of interest. This is
often done in the study of transition metal oxides where bands
bear strong orbital character~\cite{Korotin:EPJ,Amadon:prb77}. 
In the adsorbate
problem, the  band structure is far more complex, which impedes direct
band selection. We conclude that the MLWF are an optimal choice for
the problem of a large molecule on a metallic substrate.

For the substrate's MLWF, we choose to describe the $s-p$ bands only.  In order
to achieve this, the $s-p$ bands of the slab are generated
by interstitial Wannier functions, in the same way as the surface state of Cu(111)
was achieved in~\cite{Korytar:JPCM}. This is a reasonable selection,
as long as Kondo physics is considered, because the $d$ bands start at $\sim$
-3 eV below the Fermi energy.  This is away from the relevant region in energy
with an energy scale several orders of magnitude larger than the typical Kondo
scale.  Additionally, $d$-states are
rather localized, presenting small coupling with molecular states. For these
two reasons $d$ states can be safely omitted in the description of the
substrate electronic structure taking place in the Kondo effect.
The choice of the interstitial centers for the MLWF is
delicate because CuPc on Ag (100) displays a  $C_4$ symmetry \cite{Mugarza:prl}
and it is crucial to ensure that MLWF do not reduce the symmetry of the
problem. An unexpected problem of the obtention of MLWF for an extended surface
is that convergence to MLWF is more difficult as the lateral dimensions of the
super-cell are increased. The surface had to be repeatedly tested to achieve a
good description of its electronic bands within 2 eV of the Fermi energy.

\begin{figure}
\centering
  \includegraphics[height=150pt]{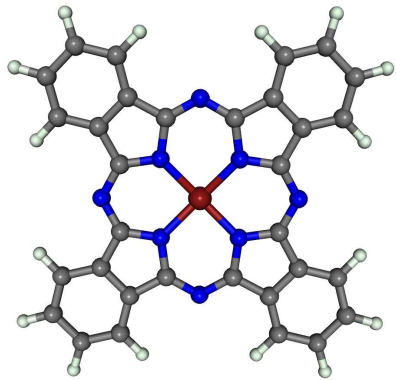}
  \caption{\label{CuPc}
 Ball-and-stick scheme of a copper phthalocyanine molecule.
 The central atom is the copper one, and the rest of atoms are nitrogen, 
 carbon and hydrogen in increasing distance from the central one.
 The  free molecule has a $D_{4h}$ symmetry
 which is reduced to $C_4$ upon adsorption on Ag (100) \cite{Mugarza:prl}.
  }
\end{figure}

The molecule's MLWF are obtained jointly with the substrate ones by using the
disentanglement method \cite{Souza:PRB}. In order to achieve the
clear partitioning between molecule and substrate, the initial set of trial
functions consists of 5 $d$ orbitals of the central copper atom (see
\fref{CuPc}), 16 $s$ states for C-H bonds, 32 $pz$ orbitals for every
carbon atom, 24 orbitals for the nitrogen atoms and 40 $s$ orbitals for C-C and
C-N bonds.  This Wannier function description was tested in the two
separate systems: the clean Ag (100) surface and the gas phase copper
phthalocyanine molecule. 

\subsubsection{\label{Sec:mo}One-particle Hamiltonian}

In the reference \cite{Korytar:JPCM}, we showed how to obtain an
Anderson-like Hamiltonian from a Kohn-Sham Hamiltonian in a MLWF basis
set. There, the evaluation of the main quantities of an atomic magnetic
impurity in a non-magnetic host was described. Especial emphasis was
put on obtaining the intra-atomic Coulomb energy. In the present
section, we extend that study to the case of molecules. The Hamiltonian
terms have to be expressed in a suitable way to be able to apply the NCA
scheme to account for Kondo physics on molecular orbitals.

Thanks to the localization and orthogonality of the MLWF basis, the
Wannier-projected Kohn-Sham Hamiltonian can be organized into four blocks
\begin{equation}
\label{Eq:KSHam}
  \left(
  \begin{array}{cc}
    H_{mol} & V' \\
    V'^\dagger & H_{slab}
  \end{array}
  \right),
\end{equation}
where $H_{mol}$ contains matrix elements between molecular Wannier functions,
the second block following the diagonal, $ H_{slab}$, refers to Wannier functions of the
silver slab, while the off-diagonal blocks, $V'$, involve couplings between molecule
and slab.

The eigenstates of $H_{mol}$ are the molecular
orbitals now obtained for the MLWF transformed Hamiltonian. 
As we shall see below, CuPc is a magnetic molecule because one electron is kept
in copper's $d_{x^2-y^2}$ orbital.
Furthermore, upon adsorption CuPc captures
one more electron in the first $\pi$-orbital, the doubly degenerated LUMO.
Hence, this open-shell structure, when hybridized with the substrate, gives rise to
the electron fluctuations of the Kondo effect. These states are singled out of the
problem and we define a special subspace for them, $\mathscr{H}_{imp}$, the
``impurity'' subspace. The rest of states is lumped together into the 
substrate's subspace, $\mathscr{H}_{subs}$.

Hence, we first apply to (\ref{Eq:KSHam}) a unitary transformation
that diagonalizes $H_{mol}$ and leaves $H_{slab}$ untouched. In this
way, we obtain the molecular orbitals of the molecular part of the Hamiltonian,
so that the first block of (\ref{Eq:KSHam}) becomes diagonal.
Explicitly,
\begin{equation}
\label{Eq:KSHam2}
  \left(
  \begin{array}{cc}
    TH_{mol}T^\dagger & TV' \\
    (TV')^\dagger & H_{slab}
  \end{array}
  \right).
\end{equation}
The unitary matrix $T$ defines the transformation from MLWF of the 
molecule to molecular orbitals.
The second step is to choose the 
molecular orbitals that play a role in the Kondo effect. These orbitals
are placed in the first rows and columns of
(\ref{Eq:KSHam2}) and define the subspace $\mathscr H_{imp}$.

After this rearrangement, the matrix of the Wannier-projected Kohn-Sham Hamiltonian 
reads
\begin{equation}
\label{Eq:KSHam3}
  \left(
  \begin{array}{c|ccc}
    H_{imp} & & V & \\
    \hline
            & &   & \\
    V^{\dagger} &   & H_{subs} & \\
            & &   &
  \end{array}
  \right)
\end{equation}
where the first block is diagonal and contains selected eigenenergies of
$H_{mol}$, the block $H_{subs}$ contains Hamiltonian matrix elements in
$\mathscr H_{subs}$ and the block $V$ contains couplings between states
in $\mathscr H_{imp}$ and $\mathscr{H}_{subs}$. The molecular orbitals
that do not directly intervene in the Kondo effect are included in the
substrate.  Hence, the full mean-field structure of the DFT calculation
is preserved within the inner window of the MLWF transformation which
is 2 eV around the Fermi energy in the present calculation.

In the present case, the obtention of an Anderson-like Hamiltonian
is facilitated by the small values of the intrinsic $U$ term in LDA.
As shown in \cite{Korytar:JPCM}, one can extract the
intrinsic $U$ term when writing the full Kohn-Sham Hamiltonian in a Wannier-basis
set. In that reference, the Co-atom impurity had values of $U$ around 1 eV.
For the present method to work, one should subtract the intrinsic $U$ term
such as is done in the LDA$+U$ method~\cite{Anisimov:PRB,Dudarev:PRB} or
in recent parameterizations of model Hamiltonians~\cite{Jacob:PRB}. Anisimov and
coworkers show that this intrinsic $U$ term is related to the Hund's rule
exchange and, thus, is much smaller than the actual Coulomb $U$ values. This
is particularly true in the case of molecules. In the case of CuPc,
we have evaluated the intrinsic $U$-term to be $\sim30$ meV, and hence
negligible in front of the molecular level values and hybridization
function.

\subsubsection{ \label{Sec:methodGamma} Ab-initio calculation of a 
hybridization function}

The hybridization function has a fundamental role in the impurity
physics \cite{Kotliar:RMP}. It is
of uttermost importance to do accurate evaluations of this function for
numerical applications. In order to achieve this, we diagonalize $H_{subs}$,
(\ref{Eq:KSHam3}), obtaining Bloch states $\ket{\ind}$ and Bloch energies
$\epsilon_\ind$, $n$ is the band
index and $\mathbf k$ is from the discretized Brillouin zone with
$50\times50\times1$ k-points. Convergency on $\mathbf k$-points was checked by
a four-fold increase of the $\mathbf k$-point sampling. 
The couplings $V$ of \eqref{Eq:KSHam3} are transformed to the 
$\ket{\ind}$ basis accordingly; we label them $V^{\phantom *}_{\ind,m}$,
where $m$ indexes states in $\mathscr H_{imp}$.
After these manipulations, we can calculate
\begin{equation}
\label{Eq:Gamma1}
\Gamma_{mm'}(\omega) = \sum_\ind
V^*_{\ind,m}V^{\phantom *}_{\ind,m'}
\delta(\omega-\epsilon_\ind)
\end{equation}
from the one-particle Hamiltonian (\ref{Eq:KSHam3}). It is
a matrix in $\mathscr H_{imp}$.


We emphasize that the Hamiltonian (\ref{Eq:KSHam3}) is obtained by projecting
the Kohn-Sham electronic structure onto a set of $N$ MLWF.  The smaller
$N$-dimensional subspace spanned by MLWF is
designed in order to represent the selected energy bands around the Fermi
level, with the same level of accuracy as the original bands of the
$M$-dimensional space of Kohn-Sham states. The total number of Wannier
functions $N$ is the sum of the MLWF of the molecule and of the substrate.
The Wannier description has been tested in separate CuPc and Ag(100) systems.
$M$ usually means the highest band
output from the \emph{ab-initio} calculation.  Interestingly, we found that it
is very important to verify the \emph{convergence} in $M$. Inclusion of bands
with energy of even tens of eV above the Fermi level is vital and preconditions
the correct projection onto the Wannier space. As a consequence, the important
value of $\Gamma_{mm}(\omega)$ at the Fermi level turns incorrect if $M$ is too
small. In the present case, convergence was attained for $N=705$ and
$M=3000$.

Formally, (\ref{Eq:KSHam3}) becomes the one-particle part of the Anderson
Hamiltonian on bringing $H_{subs}$ to a diagonal form. Although the procedure
we have just introduced is straightforward and essentially algorithmic,
it is not correct in principle, because the on-site energies of molecular
orbitals in $H_{imp}$ contain certain part of the Coulomb energy that is
included in the mean-field-like LDA framework. Furthermore, the
inadequacy of Kohn-Sham energies of orbitals lying close to the Fermi
level is well known. In our approach, $H_{imp}$ is replaced by a parameterized
model Hamiltonian with many-body interactions. This is a natural step, in 
view of the fact that we would not have to deal with Hamiltonian partitions
like in (\ref{Eq:KSHam3}) if standard \emph{ab-initio} calculations
included the relevant many-body interactions correctly.

\subsection{Impurity electronic configurations\label{config}}
In this work, the one-particle Hamiltonian $H_{imp}$ 
is replaced by a Hamiltonian $\hat h$ with many-body interactions.
This subsection discusses the physical grounds on which $\hat h$
is designed.



The experimental analysis on the Kondo features in CuPc/Ag(100) 
\cite{Mugarza:Science} indicates that 
the two-fold degenerate $e_g$ LUMO has the main role in Kondo fluctuations.
However, DFT calculations\cite{Mugarza:prl} show that the spin in
the SOMO orbital is not quenched by charge transfer from the molecule. 
Hence, the observed Kondo resonance is interpreted \cite{Mugarza:Science}
as due to a spin $S=1$ of the molecule, formed by an electron
in LUMO and one electron in SOMO. The LUMO electron is captured
from the substrate upon adsorption.

We adopt this interpretation and model the molecule as an impurity
with a two-electron $S=1$ ground state.
Furthermore, the lifetime of the SOMO will be assumed
an order of magnitude longer than the LUMO's, as will be
confirmed by \emph{ab-initio} calculation in 
\sref{Sec:HybridizationFunction}.

\begin{figure}
\centering
\includegraphics[height=100pt]{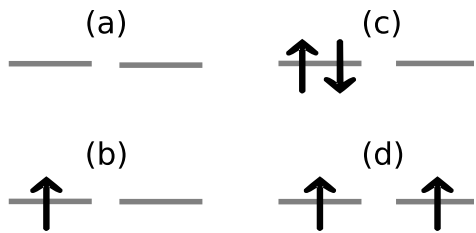}
\caption{\label{Fig:cfg} Considered electronic configurations for the 
two-fold LUMO ($e_g$):
   (a) The empty orbital (denoted by $e_g^0$),
   (b) singly occupied orbital $e_g^1$, 
   (c) doubly occupied configuration $e_g^2$,
   (d) doubly occupied configuration $e_g^{11}$. }
\end{figure}

The electron in the SOMO will be represented by a local magnetic moment
which interacts via exchange with the 
hybridized LUMO.
The multi-orbital structure of the problem
appears in the twofold degeneracy of the LUMO. The configurations that we will study
are then the ones coming from the filling of the LUMO, 
\fref{Fig:cfg}. 

We choose the relevant configurations by estimating their energetic
accessibility in the charge fluctuation process.  Then, the relevant
parameters determining the electronic configuration are: the LUMO on-site
energy $\epsilon_L$ and the charging energy $U$.  The lowest-energy
molecular configuration for the adsorbed molecule is singly occupied,
as suggested by DFT calculations \cite{Mugarza:prl}. We can estimate the
free-molecule $U$ by evaluating the energies of the neutral ($E_0$),
singly ($E_I$) and doubly ($E_{II}$) negatively charged molecule. Assuming
a simple impurity Hamiltonian, then the affinity is given by
\begin{equation} 
E_I-E_0=\epsilon_L 
\end{equation}
and the second affinity by 
\begin{equation} 
E_{II}-E_I=2\epsilon_L+U
\end{equation} 
from these equations and the total energy calculations for the
free molecular species we obtain that $U=2.08$ eV. However, $U$
is substantially screened on the metallic surface. A constrained DFT
calculation for $\pi$-systems on silver surfaces leads to a reduction of
$U$ to values between $0.5$ and  $1.0$ eV~\cite{Rohlfing}.  In the present
case we take $\epsilon_L=-0.35$ eV (see \sref{Sec:moncaResults}),
which leads to $|\epsilon_L| < U$ and  the doubly occupied configurations
$e_g^2$, $e_g^{11}$, \fref{Fig:cfg} (c,d), are higher in energy
than the neutral one, \fref{Fig:cfg} (a).  Hence, we will consider
$e_g^1-e_g^0$ fluctuations.

%

Hence, the impurity Hamiltonian contains the eigenvalues of the two LUMO as
given by the diagonalization of the MLWF Hamiltonian (\ref{Eq:KSHam2}),
plus the exchange interaction $I$ with the spin of the SOMO:
\begin{align}
\label{Eq:himp}
\hat h = \epsilon_L\sum_{a=1,2}\sum_\sigma 
\ket{a\sigma}\!\!\bra{a\sigma}
- I\mathbf S_1\cdot\mathbf S_2.
\end{align}
 We introduce Hubbard operators which automatically restrict the LUMO occupancy
to zero or one.  The sums are over spin and orbital degrees of freedom, the
latter indexed by $a$.  The second term in the Hamiltonian is the direct
exchange interaction involving the spin $\mathbf S_2$ of the SOMO and the spin
operator of the LUMO expressed through Pauli matrices $\bm\tau$ as
\begin{equation*}
\mathbf S_1 = \sum_{\sigma\sigma'}
\left(\frac{{\bm\tau}_{\sigma'\sigma}}{2}\right)\sum_{a=1,2}
\ket{a\sigma'}\!\!
\bra{a\sigma}.
\end{equation*}


If $I=0$ the Schrieffer-Wolff projection of the model onto the $e_g^1$ manifold
leads to a SU(4) Coqblin-Schrieffer model \cite{Coqblin}. 
The evolution of spectral features with
$I$ leads to a quantum phase transition of the singlet-triplet impurity
problem, that has recently received much attention experimentally \cite{Roch}
and theoretically \cite{Roura:PRB,Roura:JPCM}. Here, we will study three
cases: $I < 0$, $I=0$ and $I>0$.

\subsection{Multi-orbital non-crossing approximation\label{NCA}}

The above impurity Hamiltonian (\ref{Eq:himp}) commutes both
with the LUMO occupancy operator, $n_a = \sum_\sigma \ket{a\sigma}\!\!\bra{a\sigma}$,
and the total spin operator $\mathbf{S} = \mathbf S_1 + \mathbf S_2$ showing
that it properly describes the molecular properties.
Due to the hybridization with the substrate, the molecule will exchange electrons.
Here, we assume just fluctuations between the two above configurations, 
\fref{Fig:cfg} (a) and (b) plus the fluctuations of the SOMO spin, as
included in the definition of $\hat h$.

The full Anderson-like Hamiltonian reads 
\begin{subequations}
\label{Eq:AndersonHamiltonian}
\begin{align}
\hat{H} &=\hat h + \sum_\ks\eligand c^\dagger_\ks c^{\phantom\dagger}_\ks + \hat V\\
 \hat V &= \sum_\ks\sum_a \left(V^{\phantom*}_{\ks,a}c^\dagger_\ks \ket{0}\!\!\bra{a\sigma}
 + V_{\ks,a}^*\ket{a\sigma}\!\!\bra{0}c^{\phantom\dagger}_\ks\right).
\end{align}
\end{subequations}
The hybridization part $\hat V$ and the impurity Hamiltonian $\hat h$ are
written with Hubbard operators which change the
state of LUMO from empty to occupied or vice versa. 
Substrate electrons are described by fermionic operators $c^{\phantom\dagger}_\ks$
and energies $\eligand$. For the sake of brevity, all substrate-electronic
degrees of freedom are encapsulated in the $\mathbf k$ symbol. The band
index will be introduced when necessary.

Let
$\hat H_0 = \hat H-\hat V$.
Following Bickers \cite{Bickers:RMP}, we write down a resolvent operator of the
impurity
\begin{equation*}
\hat R(z) = \frac{1}{Z^{subs}}\sum_\Omega e^{-\beta E_\Omega}
\bra{\Omega}\frac{1}{z-(\hat H_0-E_\Omega)-\hat V}\ket{\Omega}
\end{equation*}
as a result of averaging over eigenstates $\ket{\Omega}$ of the
non-interacting substrate with
energies $E_\Omega$. The substrate partition function
is denoted by $Z^{subs}$. 
Since $\hat R(z)$ does not mix different occupancies, it can be written
as a block matrix 
\begin{equation*}
    \hat R(z) = \left(\begin{array}{cc} 
    \hat R_1(z) & 0\\ 0 & \hat R_2(z)
  \end{array}\right),
\end{equation*}
where $\hat R_1(z)$ refers to one-electron occupancy (ie LUMO
empty) and $\hat R_2(z)$ acts on two electron occupancies of the impurity.

These quantities are calculated via self-energies defined by
\begin{align*}
\hat R_1(z) &= \frac{1}{z-\hat h|_{N=1} - \hat\Sigma_1(z)} \\
\hat R_2(z) &= \frac{1}{z-\hat h|_{N=2} - \hat\Sigma_2(z)}.
\end{align*}
Following Kuramoto \cite{Kuramoto:ZPB}, we introduce basis states labeled by $\alpha$ 
for the configurations with one electron and
$\beta$ for the two-electron configurations. In the
\emph{Non-Crossing Approximation} the self-energies
for the fixed occupations are given by
all diagrams without crossings of substrate electron lines,
\begin{eqnarray*}
\Sigma_{1 \alpha'\alpha}(\omega) = \sum_{\beta\beta'}\sum_{\ks}
f(\eligand)V_{\ks}(\alpha'|\beta')\\
\times R_{2 \beta'\beta}(\omega + \eligand)V_{\ks}(\beta|\alpha)
\end{eqnarray*}
\begin{eqnarray*}
\Sigma_{2 \beta'\beta}(\omega) = \sum_{\alpha'\alpha}\sum_{\ks}
f(-\eligand) V_{\ks}(\beta'|\alpha')\\
\times R_{1 \alpha'\alpha}(\omega -\eligand)V_{\ks}(\alpha|\beta).
\end{eqnarray*}
The hybridization vertex $V_{\ks}(\alpha|\beta)$ comes when
emitting an electron from LUMO to the band state $\ks$ which is
accompanied by impurity transition from $\beta$ to the state with
label $\alpha$. Similarly, $V_{\ks}(\beta|\alpha)$ is brought
about when annihilating a one-electron state $\alpha$ of the impurity
and creating a two electron state by absorbing a substrate electron.
Explicitly,
\begin{subequations}
\label{Eq:multipletCoupling}
\begin{align}
V_{\ks}(\alpha|\beta) &= \bra{\alpha}c_\ks\hat V\ket{\beta}\\
V_{\ks}(\beta|\alpha) &= \bra{\beta}\hat Vc^\dagger_\ks\ket{\alpha}.
\end{align}
\end{subequations}
It is convenient to re-express the fixed-occupation self-energies (also called
bosonic and fermionic self-energies in slave-boson
approaches \cite{Coleman:PRB,Kroha:JPN,Roura:JPCM}) in terms of a hybridization
function that is directly related to the level broadening of the impurity
configurations
\begin{equation}
\label{Eq:Gamma}
\Gamma(\alpha'\beta|\beta'\alpha;\ \omega) = \sum_\ks
V_{\ks}(\alpha'|\beta')V_{\ks}(\beta|\alpha)
\delta(\omega - \eligand).
\end{equation}
Please, notice that we have not included $\pi$ or $2\pi$ factors as sometimes
is done in the literature. In order to evaluate lifetimes a factor $2\pi$ 
will be added to the diagonal terms of the hybridization function (\ref{Eq:Gamma}).

We can now represent the fixed-occupation self-energies in the form
\begin{subequations}
\label{Eq:nca-integral}
\begin{align}
\label{Eq:nca-integral1}
\Sigma_{1 \alpha'\alpha}(\omega) &= \sum_{\beta\beta'}\int
f(\omega')\Gamma(\alpha'\beta|\beta'\alpha; \omega')
R_{2 \beta'\beta}(\omega + \omega') \D\omega'\\
\label{Eq:nca-integral2}
\Sigma_{2 \beta\beta'}(\omega) &= \sum_{\alpha'\alpha}\int
f(-\omega') \Gamma(\alpha'\beta|\beta'\alpha; \omega')
R_{1 \alpha\alpha'}(\omega -\omega') \D\omega'.
\end{align}
\end{subequations}
This permits us to efficiently perform all computations using fast Fourier
transforms.

Let us write $\beta = (a,S,S^z)$, where $a=1,2$ indexes the two orbitals of
LUMO, $S$ the total spin and $S^z$ one of its components. Similarly,
$\alpha$ will denote the projection of the spin of SOMO on the $z$-axis, the
only degree of freedom of the one-electron configurations.

In the present case, the substrate is non-magnetic, $\eligand$ and $V_{\ks,a}$
do not depend on the electron spin, $\sigma$, and the expression for the
hybridization factorizes into spin and orbital parts (see Appendix). This
considerably simplifies the equations~\eqref{Eq:nca-integral}. Introducing
\begin{subequations}
\label{Eq:monca-res}
\begin{align}
R_{1}(\omega) &= [\omega -\Sigma_1(\omega)]^{-1}\\
\left[R^{S=0}_{2}(\omega)\right]^{-1}_{aa'} &= (\omega - \epsilon_a+\frac{3}{4}I)\delta_{aa'}
        -\Sigma_{2,aa'}(\omega)\\
\left[R^{S=1}_{2}(\omega)\right]^{-1}_{aa'} &= (\omega - \epsilon_a -\frac{1}{4}I)\delta_{aa'}
        -\Sigma_{2,aa'}(\omega)
\end{align}
\end{subequations}

and defining the orbital part of the hybridization function by (see (\ref{Gamma_orb}))
\begin{equation}
\Gamma_{aa'}(\omega) = \sum_{\mathbf k}
V^*_{\mathbf ka'}V^{\phantom *}_{\mathbf ka}
\delta(\omega - \epsilon_{\mathbf k}),
\end{equation}

we can write

\begin{subequations}
\label{Eq:monca-self}
\begin{align}
\nonumber
\Sigma_1(\omega) &= \sum_{aa'}\int f(\omega')\Gamma_{aa'}(\omega')\times\\
&\left[\frac{3}{2} R^{S=1}_{2,a'a}(\omega+\omega') +
\half R^{S=0}_{2,a'a}(\omega+\omega')
\right]\text{d}\omega'\\
\Sigma_{2,aa'}(\omega) &= \int f(-\omega')\Gamma_{aa'}(\omega')R_1(\omega-\omega')\text{d}\omega'.
\end{align}
\end{subequations}
The resolvent for the two-electron configurations $R^{S}_{2,aa'}$ now depends
on the total spin and is a matrix in the orbital space.
The equations (\ref{Eq:monca-res},\ref{Eq:monca-self}) constitute a
self-consistent system.

The main quantity of interest, the Green's function of LUMO, will be calculated
from the general time-ordered correlation function of Hubbard 
operators, defined through the thermal average
\begin{equation}
\label{Eq:gfhubbard}
G(\alpha'\beta', \tau|\beta\alpha, 0) = -T_\tau\biggl\langle e^{\tau H}
\ket{\alpha'}\!\!\bra{\beta'}e^{-\tau H}
\ \ket{\beta}\!\!\bra{\alpha}\biggr\rangle.
\end{equation}
Starting from the latter expression,
we average over the singly-occupied configurations $\alpha\alpha'$
and take the Fourier transform \cite{Kuramoto:ZPB} 
to obtain the  Green's function of LUMO for the given
spin multiplet
\begin{eqnarray}
G_{aa'}^S(\omega) = \frac{1}{Z_i}\int \bigl [
R_{2,aa'}^S(\omega+\epsilon)A_1(\epsilon)- \\ -A_{2,aa'}^S(\epsilon)
R_1(\epsilon-\omega)\bigr ]e^{-\beta\epsilon}\text{d}\epsilon.
\label{physical}
\end{eqnarray}
The quantities $A_{2,aa'}$ and $A_1$ are the spectral functions of
the resolvents and $Z_i$ is the impurity partition function.
We follow Kuramoto \cite{Kuramoto:ZPB} and use defect propagators in 
the numerical implementation of (\ref{physical}), see Appendix. 

The spectral function of LUMO is calculated by tracing over orbital
and spin degrees of freedom,
\begin{equation}
\label{Eq:SpectralFunction}
A(\omega) = -\frac{1}{\pi}\Im m \sum_{a}\left[G^{S=0}_{aa}(\omega) +
 3\ G^{S=1}_{aa}(\omega)\right].
\end{equation}

We see that the general Green's function
\eqref{Eq:gfhubbard} factorizes into channels of the total spin of 
an incoming electron (hole) and the molecule \eqref{physical}.
This fact has been used to model magnetic inelastic effects
induced by the STM \cite{Gauyacq1,Gauyacq2,Gauyacq3}. 
%
%
\section{Results: Kondo Physics of C\lowercase{u}P\lowercase{c} on A\lowercase{g} (100)\label{Results}}

Full detail on the adsorption of  CuPc on Ag (100) is given in other
publications \cite{Mugarza:prl,Roberto}. Here, we build on those results and
apply our methodology to obtain the electronic structure in presence of the
Kondo effect. 
First, we build the Anderson-like Hamiltonian following the
recipes of section~\ref{Sec:mo} and next, we solve the multi-orbital NCA
equations, section~\ref{NCA}, presenting the spectral function or PDOS onto
CuPc doubly degenerated LUMO. Special care is given to the evaluation and
presentation of the hybridization function, $\Gamma_{mm'}(\omega)$, of
section~\ref{Sec:methodGamma} because most of the relevant Kondo physics is
associated with the symmetry and values of this function. 

\subsection{Hamiltonian for CuPc on Ag (100)}

\subsubsection{Molecular orbitals}

Thanks to the partitioning made possible by the MLWF basis, we can
extract the molecular
Hamiltonian $H_{mol}$ (\ref{Eq:KSHam}) computed
from the LDA calculation of CuPc on Ag (100) \cite{Mugarza:prl,Roberto}.
Table~\ref{Tab:onsites} presents eigenenergies of $H_{mol}$ the molecular orbitals
close to the substrate's Fermi level. The orbitals are then closely related to
the gas-phase CuPc ones and hence, we denote them by SOMO, LUMO and highest occupied molecular
orbital (HOMO) labels.
For comparison, 
the results for the LDA \textsc{Siesta} calculation  of the
gas-phase molecule of the same molecular
geometry are also given. The MLWF qualitatively reproduce the 
gas phase molecular levels. 
We emphasize that the $\epsilon$ (MLWF) refer to molecular orbitals screened by
the metal but without direct hybridization and the $\epsilon$ (\textsc{Siesta})
are for a true gas-phase
molecule without any screening or coupling with an external metal.

\begin{figure}
\centering
  \includegraphics[height=150pt]{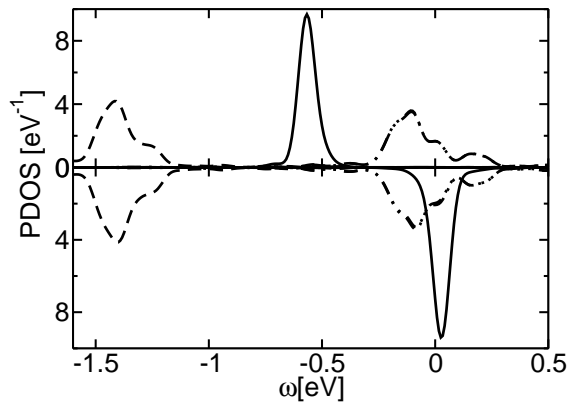}
  \caption{\label{Fig:mowpdos}
    Density of states projected onto molecular orbitals (PDOS).
    The projection onto the SOMO is given by bold lines, onto
    the HOMO by  dashed lines and onto the LUMO by
    dash-dotted lines. 
    Energies are with respect to the Fermi energy. Majority spin
    is shown in the upper panel and
    minority in the lower one. The PDOS are convoluted with
    a $50$-meV Gaussian.
  }
\end{figure}

Figure~\ref{Fig:mowpdos}  shows the PDOS onto these four molecular
orbitals.  The calculation is performed for the LDA k-point grid
($2\times2\times1$) which is clearly insufficient to account for the
continuum character of the substrate electronic states. Hence, the peaks
have been slightly broadened using a Gaussian broadening of 50 meV.
Despite the qualitative character of this figure, much information can be
gleaned from it.  The difference in widths between SOMO on one hand and
LUMO and HOMO on the other hand is substantial. The SOMO
peak basically presents the numerical width, and is a featureless peak,
while the LUMO shows oscillations and spreads over several hundreds of meV.
Then, the latter orbitals hybridize more strongly with the substrate.
From this figure we can conclude that while the LUMO width is in the range of a few hundreds of meV, the SOMO is significantly less. This is in agreement
with the move involved calculations of the hybridization function that
we present in the next section.

The PDOS are in excellent agreement with the PDOS
on molecular orbitals from DFT results \cite{Roberto}. Both the eigen-energies
and the PDOS give support to our method of transforming the DFT Hamiltonian to a
MLWF and selecting the impurity Hamiltonian using (\ref{Eq:KSHam3}).

\begin{table}[hb]
  \caption{\label{Tab:onsites}
  Eigenenergies of the molecular orbitals for the impurity Hamiltonian
  evaluated with MLWF using (\ref{Eq:KSHam2}), compared with the gas
  phase LDA calculation using \textsc{Siesta}, for the same geometry of the
  CuPc molecule.  All energies in the third column were shifted, so that the
  energies of SOMO ($\downarrow$) coincide.  Energies of the second LUMO
  state are in parenthesis.  The comparison is only indicative that the MLWF
  capture the molecular properties since the two calculations are performed
  for different systems: the MLWF refers to molecular orbitals screened by
  the metal but without direct hybridization and the \textsc{Siesta} column
  is for a true gas-phase molecule without any screening or coupling with an
  external metal.}

  \begin{indented}
  \item[]
  \begin{tabular}{ccc}
  \br
         &$\epsilon$ (MLWF) [eV] & $\epsilon$ (\textsc{Siesta}) [eV]\\
  \mr
   HOMO ($\uparrow$)  &-1.110    &-0.888    \\
   HOMO ($\downarrow$)  &-1.112  &-0.890 \\
   SOMO ($\uparrow$)  &-0.549    &-0.714     \\
   SOMO ($\downarrow$) & 0.040   & 0.040    \\
   LUMO ($\uparrow$)  & 0.185 (0.190) & 0.496 (0.498)     \\
   LUMO ($\downarrow$)  &  0.211 (0.216) & 0.520 (0.522)\\
  \br
  \end{tabular}
  \end{indented}
\end{table}

It is difficult to conclude on the actual occupancies from the PDOS.
Indeed, the PDOS numerical broadening
thwarts any precise
calculation of orbital occupancies, and the definition itself
of occupancy of a molecular orbital in a chemisorbed system is
somewhat arbitrary. 
As we show in the next section, the actual peak width of SOMO
is negligible; the SOMO then remains singly occupied as in the gas phase.
The LUMO peaks in \fref{Fig:mowpdos} cross the Fermi level,
hence these orbitals capture charge from the substrate. The LUMO are then
the only orbitals that participate in charge transfer from the surface.

\subsubsection{\label{Sec:HybridizationFunction}Hybridization function}

The lack of continuum in the PDOS calculation is cured in the calculation of
the  hybridization function (\ref{Eq:Gamma1}). Thanks to the smaller size
of the MLWF Hamiltonian (\ref{Eq:KSHam3}), we can now find the Bloch
functions for a very dense k-point grid. 

The off-diagonal parts of $\Gamma_{mm'}(\omega)$ are very small ($<1$ meV).
This is a very strong result that shows that the substrate does not mix
the different molecular orbitals among themselves.  Hence, every molecular
orbital defines an electronic channel of the system. 

\begin{figure}
\centering
  \includegraphics[height=170pt]{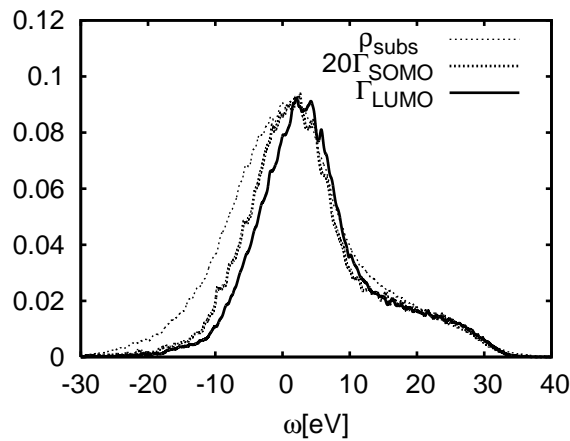}
  \caption{\label{Fig:Gamma}
    Diagonal elements of the hybridization function matrix (in eV)  as
    a function of the electron energy, $\omega$, with respect to the Fermi level.
    Only the LUMO and the SOMO (note the factor 20) are considered.
    For comparison, the
    substrate-projected density of states is given in arbitrary units.    
    The delta function of (\ref{Eq:Gamma1}) has been replaced by a
    5-meV wide Gaussian; the k-point sampling is $50\times50\times1$. 
  }
\end{figure}

The diagonal elements for SOMO and LUMO orbitals are presented in
\fref{Fig:Gamma}, along with the density of states of the substrate.  The
three curves are very similar in shape. By comparing the hybridization
functions with the substrate DOS, we notice that the hybridization function
grows more slowly than the DOS as the electron energy increases.  This is an
effect due to the couplings, $V$, between the molecular orbitals and the
substrate electronic structure. However, at higher energy, it is the DOS that
controls the behavior of the hybridization function with energy. Figure
~\ref{Fig:Gamma} shows that the SOMO width has to be multiplied by 20 to be
comparable to the LUMO one.  The SOMO orbital has then a very small mixing with
the substrate as compared to the LUMO. 
Finally,  in the $U=0$ picture, the FWHM for the LUMO is $2\pi
\Gamma_{aa}(\omega=\epsilon_{L}) = 440$ meV and is comparable to the
FWHM of the PDOS peak, \fref{Fig:mowpdos}, obtained with an insufficient
k-point sampling.

We have also evaluated the hybridization function of the LUMO
with $d$ orbitals by preparing a Wannier basis-set with $d$ orbitals.
In the region of interest here (some 2 eV about the Fermi energy),
this hybridization is strictly zero due to the lack of $d$ states at
these energies. However, when resonant with Ag $d$-band (3 eV below
the Fermi energy), the hybridization function becomes larger than
the corresponding values for the $sp$ Wannier functions. At 4 eV below the 
Fermi energy, the $d$-electron hybridization has maximum of 0.032 eV, while 
the $sp$ one is 0.035 eV. The consequence
of this is that the spectral function at -4 eV will not be just a simple
Lorentzian tail. However, the LUMO orbital is some hundreds of meV away
from the Fermi energy, and the effect of the $d$-band contribution to the
overall shape of the LUMO spectral function is negligible both for Kondo
physics and for the one-electron spectral shape.

Finally, we comment on the problem of broken spin symmetry in DFT.
Spin-polarized LDA implies two subsystems: minority and majority spin. This in
turn says that we have two sets of Wannier functions, two distinct substrates,
hybridization functions, etc. The main effect of breaking the spin invariance
in our DFT calculation is that the SOMO occupancy is $n_{SOMO}\approx 1$.  As a
secondary effect, the substrate and molecule become slightly spin-polarized as
well. However, this effect is perturbational \cite{Anderson:PR} and is not an
intrinsic property of the bare substrate and impurity of the Anderson model. In
what follows we drop the minority spin data of the hybridization function and
restore the spin symmetry.

\subsection{Multi-orbital NCA results\label{Sec:moncaResults}}

Our LDA calculations yield a hybridization function for the
SOMO, $\Gamma_S$,
ten times smaller than the one for the LUMO levels.
From these values we obtain
$\Gamma_S\approx 4.5$ meV. We use the standard expression for the Kondo
temperature \cite{Hewson:Kondo}
\begin{equation}
 T_{K,S} = D \exp \left(-\frac{|\epsilon_S|}{2\Gamma_S} \right)  \label{T_K}
\end{equation}
and take the \emph{ab-initio} value for the on-site energy from
the \tref{Tab:onsites}.
The bandwidth is 2D, where a rectangular DOS is typically assumed.
Here we have taken D as 10 eV because the DOS of Fig. 3 integrates to 703 
states, while the calculated DOS at the Fermi energy is 32.28 eV$^{-1}$.
Hence 2D=703/32.28=21.8 eV, and D turns out roughly 10 eV. 
We get $T_{K,S}$ of the range $10^{-26}$ eV. Thus, the energy scale
that would correspond to a Kondo effect on the SOMO orbital (without exchange 
coupling to LUMO) is unobservable. 

\begin{figure}
\centering
\includegraphics[height=150pt]{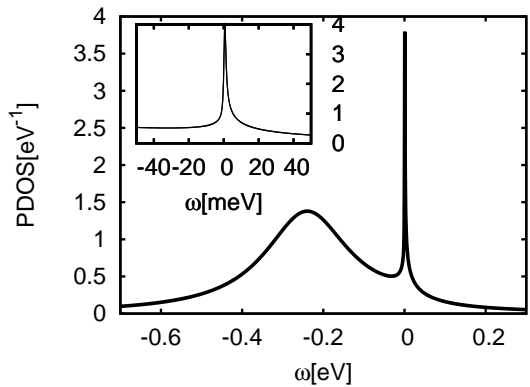}
\caption{\label{spinhalf-wide}
LUMO spectral function for the $I=0$ case. The wide peak corresponds
to a charge excitation and the narrow peak is the Kondo peak (shown
also in the inset).  }
\end{figure}

These arguments show that the Kondo coupling in this system is indeed given 
by virtual charge
fluctuations of the two-fold degenerate LUMO. Hence, we are dealing with the
impurity problem described in \sref{config},
for which we can calculate the spectral function according to the
\sref{NCA}.

However, the on-site energies of LUMO in \eqref{Eq:himp} as given
by their LDA values in the \tref{Tab:onsites} lie very close to the Fermi
level, which would correspond to a fluctuating valence regime. That has
not been observed in the experiment \cite{Mugarza:Science}. This is related to the
fact that the LDA energies of LUMO do not reflect the considerable Coulomb
repulsion in their spin splitting.
We fix these deficiencies by rescaling the $\epsilon_L/\Gamma_{aa}(\omega)$ ratio
in order to achieve a SU(4) Kondo temperature, 
$T_{K,L}=D\exp\left(-|\epsilon_L|/4\Gamma_{aa}\right)$, of $\sim$20 K as observed in the 
measurements \cite{Mugarza:Science}. The results presented in this
section are calculated with $\epsilon_L=-0.35$ eV and $\Gamma_{aa}(\epsilon_L)$
rescaled to $0.01$ eV.

Below $T_{K,L}$, the model exhibits a rich variety of physics as the value of $I$ changes.
For $I\geq 0$ but smaller than the Kondo temperature $T_{K,L}$,
the SOMO is effectively decoupled from the LUMO and we recover the case of 
two degenerated LUMO on the same footing as the
electron spin: we have an SU(4) system as described above. For negative $I$,
$|I| \lesssim T_{K,L}$, the SOMO becomes screened by a two-stage Kondo effect
at very low temperatures.
For positive and large $I\gg\Gamma_{aa},T_{K,L}$,
the Kondo physics is that of the under-screened Kondo effect, because
of the $S=1$ molecular ground state. Hence, the model will show
singular Fermi liquid
characteristics \cite{NozieresBlandin:JP,Coleman:PRL,Posazhennikova}
in the low-energy domain.
Finally, in the limit $I\rightarrow-\infty$, we obtain 
a spin zero molecule with orbital pseudo-spin, which is over-screened
by two substrate channels. The cross-over temperature is, however,
exponentially smaller than $T_{K,L}$, due to the reduced degeneracy.
Here we present NCA spectral functions in the $I=0$ case and 
in the intermediate regimes $\Gamma_{aa}>|I| > T_{K,L}$
which are dominated by inelastic spin transitions.

Figure~\ref{spinhalf-wide} shows the spectral function of LUMO
at $T=7K$ when the exchange interaction between SOMO and LUMO is turned off,
$I=0$. The ground state of $\hat h$ is fourfold degenerate in this
case. Since each of the two orbitals of LUMO defines
an independent scattering channel in the substrate (ie 
$\Gamma_{aa'}(\omega)$ is diagonal), the LUMO is subject to a $SU(4)$ Kondo effect.
The spectral function shows a broad charge-excitation peak
of the FWHM given by $4\cdot 2\pi\Gamma_{LL}(\epsilon_L)$ and a narrow
Kondo peak.

\begin{figure}
\centering
\includegraphics[height=150pt]{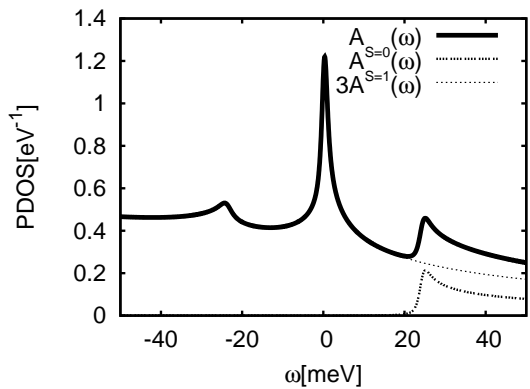}
\caption{\label{spinone}
LUMO spectral function for $I=25$ meV (bold line). The spin zero and spin one
channel contributions are presented. 
The positive-energy satellite corresponds to
injection of one electron in the excited electronic structure of the system: the
spin zero channel. The negative-energy peak corresponds to removing one electron
from the molecular ground state in the $S=1$ channel.}
\end{figure}

When the SOMO and LUMO spins are subject to a ferromagnetic interaction ($I>0$) the
ground state of the molecule without couplings to the substrate is an
orbitally degenerate $S=1$. The excited state is a $S=0$ orbital doublet.
Now, two satellites develop at $\pm I$ from
the Kondo peak, \fref{spinone}. 
These satellites are inelastic replicas of the Kondo peak since
the spin excitation energy is exactly $I$. Decomposition into spin channels
yields that the Stokes ($+I$) satellite is in the $S=0$ channel,
while the anti-Stokes peak as well as the Kondo peak are in the $S=1$ channel.
This result can be understood by recalling the meaning of the spectral function. The
spectral function at $T=0$ yields the probability density to inject one electron
in the system when $\omega>0$ or to inject a hole when $\omega<0$. 
Hence, the positive energy satellite corresponds to an excited Kondo effect
triggered by the injection of an electron to the $S=0$ state, which is an excited
state of $\hat h$.
Similar results have been found by Roura Bas and
Aligia \cite{Roura:PRB,Roura:JPCM}.

When the interaction is anti-ferromagnetic ($I<0$), the ground state of $\hat h$
is an orbital doublet and is followed by six excited states of spin one.
A noteworthy feature of the spectral function are the steps typical for
an inelastic spin-flip transition (see for example
\cite{Gauyacq1,Gauyacq2} and references therein). The inelastic
steps enhanced by Kondo effect have already been
explored in the case of singlet-triplet transitions in nanotubes \cite{Paaske}.
The spin channel analysis yields that the Stokes peak is now $S=1$ which again
can be rationalized by noticing that it corresponds to injecting
an electron in an excited $S=1$ state and the transition is enhanced by
an excited Kondo effect.
By the same token, the anti-Stokes
peak is $S=0$ since a hole is created in this channel. 

\begin{figure}
\centering
\includegraphics[height=150pt]{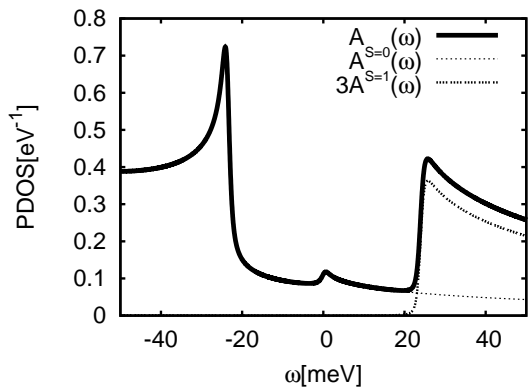}
\caption{\label{spinzero}
LUMO spectral function for $I=-25$ meV.  The spin zero and spin one
components are presented. 
The two excitation steps correspond to inelastic transitions enhanced by
the Kondo effect. As in the $S=1$ case, $I=25$ meV, the steps belong to a channel
with well defined spin. The positive-energy step is given by the spin triplet ($S=1$)
channel and the low energy peak is due to the $S=0$ channel.}
\end{figure}

The spectral function, \fref{spinzero}, shows a small peak on the
Fermi level. This peak cannot correspond to a spin-flip Kondo effect. 
The orbital Kondo resonance is ruled out by its exponentially
suppressed energy scale $\propto \exp{\left(-|\epsilon_L|/2\Gamma_{aa}\right)}$
as compared to $T_{K,L}$.
NCA is known to produce spurious peaks at the Fermi level~\cite{GreweNCA,
Kuramoto3:ZP,Bickers:RMP}. The temperature scale at which they appear
is given by~\cite{Hewson:Kondo} $T_S = T_{K,L}/5
[T_{K,L}/\Gamma_{aa}]^{5/3} = 0.18$ K (strictly valid for $I=0$ only).
All calculations 
shown here are performed at 7 K, well above the pathology temperature $T_S$.
We conjecture that the zero energy structure is related to the problems
of NCA to reproduce spin-split Kondo peaks, as detailed in 
\cite{Kroha:JPN}, p.~25.

A common feature of both $I>0$ and $I<0$ spectral
functions (figures \ref{spinone},\ref{spinzero}) is a certain asymmetry 
of the Stokes and anti-Stokes peaks. This is caused by asymmetry
of the hybridization function with respect to the Fermi energy, or by
the $U=\infty$ approximation. We have explicitly verified that the
hybridization function has very little effect on the respective heights
of both satellites. Indeed, the asymmetry
comes from the fact that system is out of the particle-hole symmetry.
Then we assign the asymmetry of the satellites
to the infinite Coulomb repulsion.

\section{Discussion}

The evaluated LDA spectral function, 
\fref{Fig:mowpdos}, seems to suggest
a fluctuating valence state while a Kondo peak have been experimentally
found~\cite{Mugarza:Science}.
Moreover,
if we added a strong Coulomb term to the LUMO, the corresponding 
Anderson Hamiltonian would correspond to an empty-impurity regime
according to the scaling theory \cite{Haldane:prl}.
For these reasons, we had to rescale the Kondo temperature in our model
by shifting the LUMO level and changing the hybridization strength
as we showed in the previous section.

These results show that LDA is unreliable to
furnish quantitative data. Any {\em ab-initio} method is bound to failure when
trying to estimate the Kondo temperature given the exponential dependence 
on the main {\em ab-initio} ingredients: the level position
and hybridization. Nevertheless, {\em ab-initio} calculations can give 
valuable qualitative input since the correct
electronic symmetry addressing both the spin and orbital channels can be
directly obtained from DFT. Our analysis of the hybridization function has
yielded important information: $(i)$ the existence of two well-defined orbital
channels originating in the hybridization of the two LUMO with the substrate
has been proved because non-diagonal terms of $\Gamma_{aa'}(\omega)$ are vanishingly
small, $(ii)$ both LUMO and SOMO are partially charged in the adsorbed system,
the LUMO due the sizeable values of $\Gamma_{aa}$ that leads to charge
transfer from the substrate, and the SOMO for the vanishing value of
$\Gamma_{S}$ that leaves its spin unperturbed. Hence, LDA reveals
the orbital $SU(2)$ symmetry associated with
LUMO electron and its substrate channels.

The failure of the present LDA calculations to yield a realistic
$\epsilon_L/\Gamma_{aa}$ ratio can be traced back to the complete failure in giving
the qualitative Kondo physics of the CuPc/Ag(100) system. As we just saw, 
the PDOS onto Kohn-Sham states
predicts the system to be in a mixed-valence regime while the experimental data
show it is a Kondo system. This points out at the present failures, namely, the
LUMO occupation is poorly accounted for in LDA. Previous works have claimed
that LDA is not capable to yield correct hybridization functions,
because of the lack of charge discontinuity in the
exchange-and-correlation functional \cite{Evers}. While this is clearly the
case in transport calculations, the present hybridization function describes a
static situation, where the strength of the coupling to the substrate is
evaluated. Furthermore, the hybridization function evaluates the transparency
of the barrier between the molecule and the substrate \cite{Gauyacq4}.
Hence, we do not think
that the hybridization function is affected by the charge discontinuity
problem. We think it is rather the electronic configuration that is affected by
the charge discontinuity problem yielding wrong occupancies, the alignment of
the molecular levels and finally the wrong qualitative picture.

Finally, the cure has already been advanced in the literature by using LDA$+U$
methods \cite{Neaton}. As in the reference~\cite{Neaton}, $U$ has to be computed
for the LUMO orbitals as has been recently realized by an increasing number of
groups \cite{Rohlfing,Neaton}. 

It is for these failures of the LDA calculations that the 
$\epsilon_L/\Gamma_{aa}$
had to be adapted in order to obtain reasonable spectral functions in
the \sref{Sec:moncaResults}. In the physically relevant case of the positive
SOMO-LUMO exchange coupling of $I=25$ meV we obtained the experimentally
observed three-peak structure in the vicinity of the Fermi level \cite{Mugarza:Science}.



In spite of the shortcomings stemming from LDA's description of the LUMO
shown in this work,
we emphasize that the Wannier-based approach is
not restricted to the use of LDA functional and can be interfaced to
an arbitrary \emph{ab-initio} method which provides a one-particle 
structure as for example the promising \emph{GW+MLWF} method \cite{HamannVan:prb}.
Moreover, Wannier functions allow to take a step beyond the super-cell
approach and simulate a true impurity problem (ie a single molecule
on a surface) by reconstructing a semi-infinite substrate out of the
tight-binding Hamiltonian elements obtained in a super-cell.
\section{Conclusion}

We have implemented a multi-orbital non-crossing approximation approach based
on a standard one-electron \emph{ab-initio} electronic structure.
The Kohn-Sham orbitals of a DFT calculation are transformed to a 
maximally localized Wannier function (MLWF) basis
set such that a tight-binding like Hamiltonian is obtained, in view of the locality
and orthogonality of MLWF. This procedure is in principle algorithmic and
permits us to have a quantum impurity model from a DFT calculation.

We have applied this methodology to the case of a copper phthalocyanine (CuPc)
molecule adsorbed on Ag(100) following existing experimental
work \cite{Mugarza:prl,Mugarza:Science}. From our DFT calculations we conclude
that the two-fold degenerate LUMO captures charge and this can give rise
to the Kondo effect. The spin of the SOMO is a spectator because
its Kondo energy scale is many orders of magnitude smaller than the
one of the LUMO. However, intramolecular exchange interaction between the SOMO
and LUMO spin gives rise to a rich singlet-triplet phenomenology that our
numerical procedure captures. Our calculations show that two well defined
orbital channels emerge in the substrate as dictated by the $C_4$ point group 
symmetry. We have further investigated the impurity spectral function in terms of
spin channels and we have rationalized the inelastic features of the spectral
function for both (spin) singlet and triplet ground states. 

This physics has been obtained by rescaling the computed Kondo temperature to
fit the experimental one. Indeed, our LDA-based calculation fails to yield a
Kondo ground state and rather predicts a fluctuating-valence system. We attribute
this error to the lack of charge discontinuity in the LDA
exchange-and-correlation functional and suggest that alternative LDA$+U$ methods
for the one-electron calculation will improve the agreement with the
experimental electronic structure of CuPc on Ag(100).

\appendix
\section{Non-crossing approximation}

\subsection{Spin coefficients for a singlet-triplet impurity}
The couplings between $\alpha$ and $\beta$ (see \eqref{Eq:multipletCoupling})
are given by the expression
\begin{align*}
V_\ks(\alpha'|\beta) &= V_{\ks,a}\braket{\sigma\alpha}{SS^z}
\end{align*}
with $V_{\ks,a}$ from the Hamiltonian \eqref{Eq:AndersonHamiltonian}
and the braket is a Clebsch-Gordan coefficient.
When the substrate is non-magnetic, $\eligand$ and $V_{\ks,a}$ do not
depend on $\sigma$ and the expression for hybridization intensity 
\eqref{Eq:Gamma} factorizes 
\begin{align*}
\Gamma(\alpha'a'S'{S^z}'|aSS^z\alpha;\ \omega) = \Gamma_{aa'}(\omega)
\gamma_{\alpha'\alpha}(S'{S^z}'|SS^z)
\end{align*}
into orbital
\begin{equation}
\Gamma_{aa'}(\omega) = \sum_{\mathbf k}
V^*_{\mathbf ka'}V^{\phantom *}_{\mathbf ka}
\delta(\omega - \epsilon_{\mathbf k})
\label{Gamma_orb}
\end{equation}
and spin parts
\begin{equation}
\label{Eq:gammadef}
\gamma_{\alpha\alpha'}(S'{S^z}'|SS^z)=
\sum_\sigma\braket{S'{S^z}'}{\sigma\alpha'}\!\!\braket{\sigma\alpha}{S{S^z}}.
\end{equation}
The subsequent identities have proven significant
\begin{align}
\sum_{\alpha}\gamma_{\alpha\alpha}(S'{S^z}'|SS^z) &=
\delta_{SS'}\delta_{S^z{S^z}'}\\
\sum_{S^z}\gamma_{\alpha\alpha'}(1S^z|1S^z) &= \frac{3}{2}\delta_{\alpha\alpha'}\\
\gamma_{\alpha\alpha'}(00|00) &= \frac{1}{2}\delta_{\alpha\alpha'}.
\end{align}
The first one is due to completeness (\ref{Eq:gammadef}), the second
and third can be proven using Wigner \emph{3jm} symbols \cite{Landau:QM}.

With these identities it is easy to show that the following property
holds: Let $\hat R_1^{test}(\omega)$ and $\hat R_2^{test}(\omega)$
be some functions diagonal in the total spin representation, so will the
self-energies calculated from NCA \eqref{Eq:nca-integral1},
\eqref{Eq:nca-integral2},
\begin{eqnarray*}
\Sigma^{test}_{1 \alpha'\alpha}(\omega) = \sum_{\beta\beta'}\int
f(\omega')\Gamma(\alpha'\beta|\beta'\alpha; \omega') \\
\times R^{test}_{2 \beta'\beta}(\omega + \omega') \D\omega'
\end{eqnarray*}
\begin{eqnarray*}
\Sigma^{test}_{2 \beta\beta'}(\omega) = \sum_{\alpha'\alpha}\int
f(-\omega') \Gamma(\alpha'\beta|\beta'\alpha; \omega')\\
\times R^{test}_{1 \alpha\alpha'}(\omega -\omega') \D\omega'.
\end{eqnarray*}
Since we solve the equations of NCA iteratively, starting with
resolvents having zero self-energies, we conclude that 
spin off-diagonal terms of $\hat R_{1,2}$ vanish.

\subsection{Evaluation of the physical Green's function using defect propagators}
\Eref{physical} has to be expressed in terms of defect propagators.
For convenience we introduce boldface notation for matrices in
the orbital space of LUMO, ie $\mathbf R_2^S, \mathbf A_2^S,\mathbf \Gamma,
\mathbf G^S$ for the resolvent of the two-electron configuration, its
spectral density, hybridization function \eqref{Gamma_orb}
and the Green's function of LUMO.
The starting expression (\ref{physical}) reads
\begin{eqnarray}
\label{Eq:physicalBf}
\mathbf G^S(\omega) = \frac{1}{Z_i}\int\bigl[
\mathbf R_2^S(\omega+\epsilon)A_1(\epsilon)-\\ 
-\mathbf A_2^S(\epsilon)
R_1(\epsilon-\omega)\bigr]e^{-\beta\epsilon}\text{d}\epsilon.
\end{eqnarray}
We introduce the operator $\mathfrak P$, whose effect on an arbitrary
matrix function $\mathbf X(\omega)$ is given by
\begin{equation*}
\mathfrak P\mathbf X(\omega) = \frac{i}{2\pi Z_i} e^{-\beta\omega}\left[\mathbf X(\omega)-
\mathbf X^\dagger(\omega)\right].
\end{equation*}
By applying $\mathfrak P$ on both
sides of NCA equations \eqref{Eq:monca-self} yields a self-consistent
system
\begin{eqnarray*}
\mathbf a_2^S(\omega) = \mathbf R_2^S(\omega)\int f(\omega')\mathbf\Gamma
(\omega')
a_1(\omega-\omega')\text{d}\omega'\mathbf R_2^{S\dagger}(\omega)
\end{eqnarray*}
\begin{eqnarray*}
a_1(\omega) = |A_1(\omega)|^2\half\sum_{S=0,1}(2S+1)\\
\times\int f(-\omega')
\text{Tr}\left\{\mathbf\Gamma(\omega')\mathbf a_2^{S}(\omega+\omega')\right\}
\text{d}\omega'
\end{eqnarray*}
for the defect propagators~\cite{Kuramoto3:ZP}, defined by
\begin{equation*}
a_1(\omega) = \mathfrak{P}A_1(\omega)\qquad\text{and}\qquad
\mathbf a_2(\omega) = \mathfrak{P}\mathbf A_2(\omega).
\end{equation*}
When these equations are solved, the Green's function (\ref{Eq:physicalBf})
can be expressed as
\begin{eqnarray}
\label{Eq:GfThroughDefectp}
\mathbf G^S(\omega) = \int\bigl[
\mathbf R_2^S(\omega+\epsilon)a_1(\epsilon)-\\
-\mathbf a_2^S(\epsilon)
R_1(\epsilon-\omega)\bigr]\text{d}\epsilon.
\end{eqnarray}

\ack
We are grateful to
A. Mugarza, P. Gambardella for fruitful discussions and
 providing their experimental data prior to
publication. In particular, we thank
A. Mugarza for showing us that CuPc on Ag (100) was in a triplet state.
Discussions with Jean-Pierre Gauyacq, Roberto Robles and Enric Canadell
are also gratefully acknowledged. Financial support from the Spanish
MICINN (FIS2009-12721-C04-01) is acknowledged. R.K. is supported by the
CSIC-JAE predoctoral fellowship.

\section*{References}
\bibliographystyle{unsrt}
\bibliography{references}
\end{document}